\documentclass[final,12pt,onecolumn]{IEEEtran}

\usepackage{amsmath}
\usepackage{url}
\usepackage[pdftex]{graphicx}
\usepackage{cite}
\usepackage[english]{babel}
\usepackage{color}

\renewcommand{\baselinestretch}{1.2}
\begin{document}
\title{Hierarchical Clustering for Smart Meter Electricity Loads based on Quantile Autocovariances}
%
%
%

\author{Andr\'es~M.~Alonso,
	F. Javier~Nogales
        and~Carlos~Ruiz
\thanks{A. M. Alonso, F. J. Nogales  and C. Ruiz are with the Department
of Statistics and UC3M-BS Institute for Financial Big Data (IFiBiD), University Carlos III de Madrid, Avda. de
la Universidad 30, 28911-Legan\'es, Spain. (e-mails: andres.alonso@uc3m.es; fcojavier.nogales@uc3m.es; carlos.ruiz@uc3m.es)}
}


\maketitle

\begin{abstract}
In order to improve the efficiency and sustainability of electricity systems, most countries worldwide are deploying advanced metering infrastructures, and in particular household smart meters, in the residential sector. This technology is able to record electricity load time series at a very high frequency rates, information that can be exploited to develop new clustering models to group individual households by similar consumptions patterns.
To this end, in this work we propose three hierarchical clustering methodologies that allow capturing different characteristics of the time series. These are based on a set of ``dissimilarity'' measures computed over different features: quantile auto-covariances, and simple and partial autocorrelations.
The main advantage is that they allow summarizing each time series in a few representative features so that they are computationally efficient, robust against outliers, easy to automatize, and scalable to hundreds of thousands of smart meters series.
We evaluate the performance of each clustering model in a real-world smart meter dataset with thousands of half-hourly time series. The results show how the obtained clusters identify relevant consumption behaviors of households and capture part of their geo-demographic segmentation. Moreover, we apply a supervised classification procedure to explore which features are more relevant to define each cluster.

\end{abstract}

\begin{IEEEkeywords}
	Quantile autovariances, massive time series, hierarchical clustering, smart meters.
\end{IEEEkeywords}

\IEEEpeerreviewmaketitle

\section{Introduction}
\subsection{Background and Aim}
\IEEEPARstart{M}{oved} by the need of improving the efficiency and sustainability of aging electrical systems, many countries worldwide are adopting new information and communication technologies, with special emphasis on the residential sector \cite{hierzinger2012european}.
These technologies imply a new paradigm in the economical and technical operation of distribution networks, and create new business opportunities for all the companies that take part in the electricity supply chain.

It is very relevant the extended integration of advanced metering infrastructures (AMI) \cite{depuru2011smart} with an special role played by households ``smart meters''. These devices allow recording electricity consumption data at a very high frequency rate and instantly transmit this information to the retailing and/or distribution companies.

Furthermore, as many electricity markets worldwide are open to competition in both the generation and retailing sectors, there is a growing interest by the electrical companies in using these data to increase their profit, their market share or the consumers' welfare. In this vein, the treatment of these new datasets require the research and implementation of novel  data science techniques, with practical applications on energy fraud detention, outliers identification, consumers profiling, demand response, tariff design, load forecasting, etc. \cite{yildiz2017recent}.

The special characteristics of the data stored by smart meters (hundreds of thousands, or even millions, of high frequency time series), and their combination with exogenous variables (meteorological, calendar, economical, etc.), open the possibility of designing specific clustering models for household consumers. Furthermore, these models can help to better understand the behavior of both aggregated and disaggregated electrical loads  \cite{wang2018review}, and how this knowledge can be exploited to improve electrical system's.

In particular, clustering households with similar consumption patterns has many potential applications. Retailing companies can be interested on grouping clients by consumption profiles to offer tailor-made tariffs. This may increase consumers utility while ensuring revenue-adequacy for the company. Moreover, clustering may help to identify the best candidate group of consumers to implement demand response policies. In this vein, system operators and distribution companies can benefit from clustering techniques to improve their load forecasting accuracy \cite{al2017k}, with a direct impact on system reliability or predictive maintenance.

In this work we propose different hierarchical-based clustering strategies based on a set of ``dissimilarity'' measures: quantile auto-covariance, and simple and partial autocorrelations. These strategies summarize each consumption time series in only a few representative features so that they are highly efficient, easy to automatize and scalable to hundreds of thousands of series, i.e., can be successfully implemented in large-scale applications that make use of smart meters datasets. We test the performance of our clustering models by using a real-world dataset with  thousands of electricity consumption time series. The results are promising as the obtained clusters not only identify relevant consumption patterns but also capture part of the geo-demographic segmentation of the consumers. Furthermore, we implement a multiclass supervised classification algorithm, based on decision trees, in order to characterize the most important features conditioning each cluster.

\subsection{Literature Review}\label{sec:literature}
A review of several clustering techniques to group similar electricity consumers is presented in \cite{chicco2012overview}. It is shown that the overall performance of the different techniques is related to their ability to isolate outliers.
Reference \cite{koivisto2012clustering} proposes a clustering method for household consumers based on K-means and Principal Components Analysis (PCA). The resulting clusters are subject to a multiple regression analysis to identify relevant explanatory variables.
The work in \cite{kwac2014household} addresses the consumer segmentation problem by normalizing the daily load shapes for each consumer, together with their total consumption, to apply an adaptive K-means algorithm.
A clustering model based on K-means is proposed in \cite{lavin2015clustering} to, focusing on commercial and industrial electricity consumers, identify candidate users for energy efficiency policies and their businesses opening and closing hours.
Reference \cite{mcloughlin2015clustering} evaluates and compare three clustering techniques for smart meter data: k-medoid, K-means and Self Organizing Maps (SOM), to show that the latter presented to overall best performance.
Traditional time series methods are applied in \cite{tureczek2018electricity}, like wavelets or autocorrelation analysis, to the raw smart meter data to enrich the input of a K-mean based clustering algorithm for consumers segmentation.
Reference \cite{wang2016clustering} proposes to use dynamic information, in terms of transitions between adjacent time periods, for consumers segmentation. The resulting clusters are used to evaluate their potential for demand response policies.

Several works seek to identify relevant features that condition the dynamic patterns of electricity consumers. For instance, a supervised ML model is proposed in \cite{beckel2014revealing} based on individual household consumption time series. With the same aim, \cite{kavousian2013determinants} proposed a methodology to examine smart meter data and identify important determinants of consumers electricity load.
To extend the number of features that can potentially be used for profiling consumers, \cite{gouveia2016unraveling} complement the smart meter data with door-to-door question surveys. It is shown how these new dataset improves the performance of a Ward's hierarchical clustering algorithm.
A detailed analysis of household consumption data is presented in \cite{haben2015analysis} to identify those time periods from which relevant consumption features can be extracted. Based on these features, a mixture-based clustering algorithm is proposed and evaluated by bootstrap techniques. Another mixture model framework, based on linear Gaussian approximations, is used by \cite{stephen2013enhanced} to derive relevant load profiles from individual consumption patterns.

To improve computational performance, \cite{al2017multi} presents a two-level clustering methodology to derive representative consumptions profiles based on K-means. The first level is used to obtain local profiles that are generalized in the second level.
With a similar aim, reference \cite{al2016feature} proposes a feature construction model for time series to cluster similar consumers. The model reduces the dimensionality of the problem by using conditional filters and profile errors.
An efficient frequency domain hierarchical clustering model is proposed \cite{zhong2014hierarchical} to derive adequate load profiles.
Moreover, \cite{granell2014impacts} studies how the temporal resolution of the consumption time series may have an strong impact on both the quality and computational performance of the clustering techniques.

Clustering techniques has been used also to improve the accuracy of forecasting models. In this vein, a K-means based algorithm is employed in \cite{al2017k} to derive consumption estimates and impute missing data.
The cross-similarities between consumptions series is used by \cite{bandara2017forecasting} to enhance the performance of a forecasting model, based on Long Short-term Memory (LSTM) networks.
Similarly, \cite{quilumba2014using}, implements a K-means based clustering algorithm to group similar consumers and then adjust a Neural Network (NN) forecasting model for aggregated loads. 
Another clustering K-Means based algorithm is employed in \cite{chaouch2013clustering} to household load curves to group similar consumers and enhance the performance of a nonparametric functional wavelet-kernel approach.
Reference \cite{yildiz2018household} also makes used of consumers segmentation through PCA and K-means clustering to identify typical daily consumption profiles that can improve the accuracy of a ML forecasting tool.


\subsection{Contributions}
We build part of our research on the original methodology presented in \cite{lafuente2016clustering}, which proposes to cluster time series based on quantile autocovariances distances. An extensive simulation analysis and a real-world application on daily financial time series show the ability of this approach to identify different dependence models among the series.

In the present work, and by the first time to the authors knowledge, we adapt and extend part of the methodology in \cite{lafuente2016clustering} to identify relevant clusters from massive and high frequency smart meters time series.


In particular, by considering the state of the art presented in Section \ref{sec:literature},
the main contributions of this work are fivefold:
\begin{enumerate}
	\item To summarize each smart meter time series in an small set of meaningful features: autocorrelation coefficients, partial autocorrelation coefficients and quantile autocovariances.
	\item To propose three hierarchical clustering models, based on Euclidean dissimilarity measures, computed over the previous features. The models  are computationally efficient and robust against outlier observations.
	\item To test the proposed methodology in a real dataset, including thousands of half-hourly load time series, to characterize relevant electricity consumption profiles.
	\item To make use of a supervised classification procedure (decision trees) to identify those variables (features) that have been more relevant to form the resulting clusters. 
	\item To verify that the resulting clusters are able to capture, up to some extend, the geo-demographic segmentation of household consumers.
\end{enumerate}

\subsection{Paper Organization}
This paper is organized as follows. In Section \ref{sec:Model} the proposed hierarchical clustering methodology for smart meter time series is presented. The numerical results, based on a real-world dataset are presented in Section \ref{sec:Results}. Finally, Section \ref{sec:Conclusions} presents the main conclusions derived from this work.

\section{The Clustering Methodology}\label{sec:Model}

Let's assume that we observe $N$ time series, $\{\pmb{X}_1, \pmb{X}_2, \dots, \pmb{X}_N\}$ where $\pmb{X}_i = (X_{i,t_i}, X_{i,t_i+1},\dots, X_{i,T_i})$
and $(t_i,T_i)$ denotes the first and the last times where the $i$-th time series is observed, respectively. In our dataset, the
$(t_i,T_i)$ are the same for all time series but our procedures do not require this condition since they are based on extracted
features from the time series. As mentioned in the previous section, there are many interesting features to consider as ``clustering'' variables instead of using raw data. In our case, we consider three sets of features that capture different aspects of the time series dynamic behaviour:
\begin{itemize}
	\item The set of {\it autocorrelation coefficients} of orders $(1,2, \dots, K)$, that is, we calculate the correlation coefficient
	between the variables $X_{i,t}$ and $X_{i,t+j}$ for $j = 1, 2, \dots, K$ defined by
	\begin{equation}\label{corr}
	\rho_i(t,t+j)=\frac{Cov\left(X_{i,t},X_{i,t+j}\right)}{(Var(X_{i,t}) Var(X_{i,t+j}))^{1/2}}.
	\end{equation}
	\item The set of {\it partial autocorrelation coefficients} of orders $(1,2, \dots, K)$, that is, we calculate the correlation coefficient
	between observations separated by $j$ periods, $X_{i,t}$ and $X_{i,t+j}$, when we eliminate the linear dependence due to intermediate
	values. The partial autocorrelation coefficient will be denoted by $\pi_i(t,t+j)$.
	\item The set of {\it quantile autocovariances} of order $j$ at quantile levels $(\tau, \tau') \in [0,1]^2$ defined by
	\begin{align}\label{qac}
	&\gamma_{i,(\tau, \tau')}(t,t+j)= Cov\left( I(X_{i,t}\le q_{\tau,i}), I(X_{i,t+j} \le q_{\tau',i})\right),
	\end{align}
	where $I(\cdot)$ denotes the indicator function and $q_{\tau,i}$ and $q_{\tau',i}$ are the $\tau-$ and $\tau'-$quantiles of $X_{i,t}$ and $X_{i,t+j}$,
	respectively.
\end{itemize}

It is interesting to realize the differences among features (\ref{corr}) and (\ref{qac}) since both involve the calculation of a covariance between
observations separated by $j$ periods. In (\ref{corr}), the covariance term is estimated by
\begin{align*}
&\frac{1}{T_i - j} \sum_{t=t_i}^{T_i-j} X_{i,t} X_{i,t+j} - \ \frac{1}{T_i - j} \sum_{t=t_i}^{T_i-j} X_{i,t} * \frac{1}{T_i - j} \sum_{t=t_i}^{T_i-j} X_{i,t+j},
\end{align*}
which involves the products $X_{i,t} X_{i,t+j}$ that can be distorted by extreme or outlier observations. For example, two very high loads observed at a distance of $j$ periods would spuriously increase the correlation at the $j-$lag. On the other hand, the quantile autocovariance (\ref{qac}) is estimated
by
\begin{align}\label{qac2}
    &\widehat{\gamma}_{i,(\tau, \tau')}(t,t+j)= \frac{1}{T_i - j} \sum_{t=t_i}^{T_i-j} I(X_{i,t}\le \widehat{q}_{\tau,i}) I(X_{i,t+j} \le \widehat{q}_{\tau',i}) \ - \ \tau \tau'.
\end{align}
The involved products $I(X_{i,t}\le \widehat{q}_{\tau,i}) I(X_{i,t+j} \le \widehat{q}_{\tau',i}$ are bounded which imply a negligible effect of outliers.
The expression (\ref{qac2}) can be interpreted as a mean of the number of times that values at $t$ below $\widehat{q}_{\tau,i}$ coincide with values at $t+j$ below $\widehat{q}_{\tau',i}$. The term $\tau \tau'$ is the number of coincidences that occur completely randomly. Therefore, a positive $\widehat{\gamma}_{i,(\tau, \tau')}$ means that the number of matches is greater (smaller) than expected by chance.

It should be noticed that the above characteristics, in general, depend on $t$ and $j$, but if the time series are stationary, then they
do not depend on $t$, which simplifies their analysis. For this reason, we consider the (daily) seasonal difference of the smart meter
load (logarithmic transformed) time series. That is, as the time series that will be used in this paper present an half-hourly frequency, then $X_{i,t} = \ell_{i,t} - \ell_{i,t-48}$ are the series to be clustered, where $\ell_{i,t} = \log L_{i,t}$ denotes the logarithm of the load
time series of the $i$-th smart meter. We should fix the largest lag, $K$, in the sets of autocorrelation and partial autocorrelation coefficients. We can fit autoregressive models to all the univariate time series, selecting the order by the BIC criterion, and take $K = \max_{1\leq i\leq N}(p_{i})$, where $p_{i}$ is the selected order for $i$-th time series. It is shown in \cite{alonsopenya} that this procedure provides an upper bound of the memory of $N$ stationary linear time series. The selected $K$ was 96. This selection allows us to captures the main linear dependencies in all time
series. Also, for the set of quantile autocovariances, we should fix the lag and quantile levels. In this case, following the suggestions of
\cite{lafuente2016clustering}, we use $j = 1$ and $\tau \in \{0.1, 0.5, 0.9\}$ since these values have shown that they are capable of capturing and
differentiating different types of nonlinearities. Finally, the three clustering analyzes will be based on the following sets of features:
\begin{enumerate}
	\item[a)] $\pmb{\rho}_i = \{\rho_i(1),\rho_i(2),\dots,\rho_i(96)\}_{i \in \{1,2,\dots,N\}}$
	\item[b)] $\pmb{\pi}_i = \{\pi_i(1),\pi_i(2),\dots,\pi_i(96)\}_{i \in \{1,2,\dots,N\}}$
	\item[c)]  $\pmb{\gamma}_i = \{\gamma_{i},(0.1, 0.1)$, $\gamma_{i},(0.1, 0.5)$, $\gamma_{i},(0.1, 0.9)$, $\gamma_{i}(0.5, 0.1)$, $\gamma_{i}(0.5, 0.5)$, $\gamma_{i}(0.5, 0.9)$,
	$\gamma_{i}(0.9, 0.1)$, $\gamma_{i}(0.9, 0.5)$, $\gamma_{i}(0.9, 0.9)\}_{i \in \{1,2,\dots,N\}}$
\end{enumerate}
Thus, the analysis will be based on $96 \times 1$
vectors of features for autocorrelation and partial autocorrelation coefficients and based on $9 \times 1$ vectors of features for quantile
autocovariances.
Once we have the vectors of features, we define a dissimilarity measure between time series $\pmb{X}_i$ and $\pmb{X}_j$ by the Euclidean distance of
the corresponding vectors. That is:
\begin{enumerate}
	\item[a)] $d_{AC}(\pmb{X}_i, \pmb{X}_j) = \|\pmb{\rho}_i - \pmb{\rho}_j\|_2$
	\item[b)] $d_{PAC}(\pmb{X}_i, \pmb{X}_j) = \|\pmb{\pi}_i - \pmb{\pi}_j\|_2$
	\item[c)] $d_{QC}(\pmb{X}_i, \pmb{X}_j) = \|\pmb{\gamma}_i - \pmb{\gamma}_j\|_2$
\end{enumerate}
where $\|\cdot\|$ denotes de Euclidean distance.

The distances $d_{M}(\pmb{X}_i, \pmb{X}_j)$ will be obtained for all pairs $(i,j)$ with $i\neq j$ to construct the following $N \times N$ dissimilarity matrix
\begin{align}\label{dissmatrix2}
	&\pmb{DM}_{M}=\begin{pmatrix}
	0 & d_{M}(X_{1},X_{2}) & \ldots & d_{M}(X_{1},X_{N})\\
	d_{M}(X_{2},X_{1}) & 0 & \ldots & d_{M}(X_{2},X_{N})\\
	\vdots & \vdots & \ddots & \vdots\\
	d_{M}(X_{N},X_{1}) & d_{M}(X_{N},X_{2}) & \ldots & 0
	\end{pmatrix}
\end{align}
where $M \in \{AC, PAC, QC\}$.
The dissimilarity matrix (\ref{dissmatrix2}) can be used in any cluster procedure which requires this kind of input. In particular, we
can apply hierarchical clustering since it allows us to identify clusters as well as hierarchies among the clusters. In hierarchical cluster procedures,
to decide which groups should be combined, it is necessary to choose a measure of dissimilarity (linkage criterion) between sets. It is important to
emphasize that this choice will influence the shape of the groups, since some sets could be close according to one distance and far according to another.
The three best known measures are minimum or single-linkage ($d_s$), maximum or complete-linkage ($d_c$) and average linkage ($d_a$) defined by:
\[d_s(A,B) = \min\{d(X_i,X_j): i \in A, j \in B\}\]
\[d_c(A,B) = \max\{d(X_i,X_j): i \in A, j \in B\}\]
\[d_a(A,B) = \frac{1}{n_A n_B} \sum_{i=1}^{n_A} \sum_{i=1}^{n_B} d(X_i,X_j),\]
where $A$ and $B$ are two sets of observations having $n_A$ and $n_B$ elements, respectively.

In this work, we prefer to use complete linkage as it ensures that the observations in a group are “similar” to all observations of the same group in the sense that once the cut-off point in the dendrogram has been set all the distances within of a cluster are smaller than this cut-off point.

Once we obtain the groups, an interesting question is to know which variables have been the most relevant to form these groups. This question can be addressed through 
the use of a supervised classification procedure where the labels of the observations will be the result of the clustering methodology. That is, if we have $k$ clusters, 
we will assign the labels $\{1, 2,\ldots, k\}$ to the observations of the respective clusters. These labels and the features will be the input of the supervised 
classification procedure. In this work, we will use decision trees \cite{breiman1984classification} for multiclass classification problem since for this procedure unbiased estimates 
of the predictor (feature) importance \cite{ishwaran2007} are available.

\section{Numerical Results}\label{sec:Results}
In this section we use the public energy consumption dataset from \cite{dataLondon}. It includes a sample of 5,567 households of London with their individual electricity consumption time series during 2013, in kWh (per half hour), date and time, and CACI ACORN segmentation (6 geo-demographic categories) \cite{CACI}. In particular, to validate this work's clustering methodology, we will compare the resulting clusters with the geo-demographic aggregated categories coded as ``ACORN\_GROUPED'', which classify households into three main groups: ``Affluent'', ``Comfortable''  and ``Adversity''. Moreover, the dataset is also divided into two subgroups of consumers:
\begin{itemize}
	\item[i)] \textit{std tariff}: Consumers whose electricity tariff is fixed (standard) to a constant price during the time of the study.
	\item[ii)]\textit{tou tariff}: Consumers with ``time of use'' tariff for which the electricity price is different for each hour.
\end{itemize}

In order to better characterized the inherent consumption behavior of individual households, we have focused the following study on the std tariff consumers, as these are not influenced by a variable price signal. This initial group includes approximately 4500 time series from which some of them are discarded, due a high proportion of missing observation, rendering a final subsample of around 3200 time series (households).

\begin{figure}
	\centering
	\includegraphics[width=4.5in]{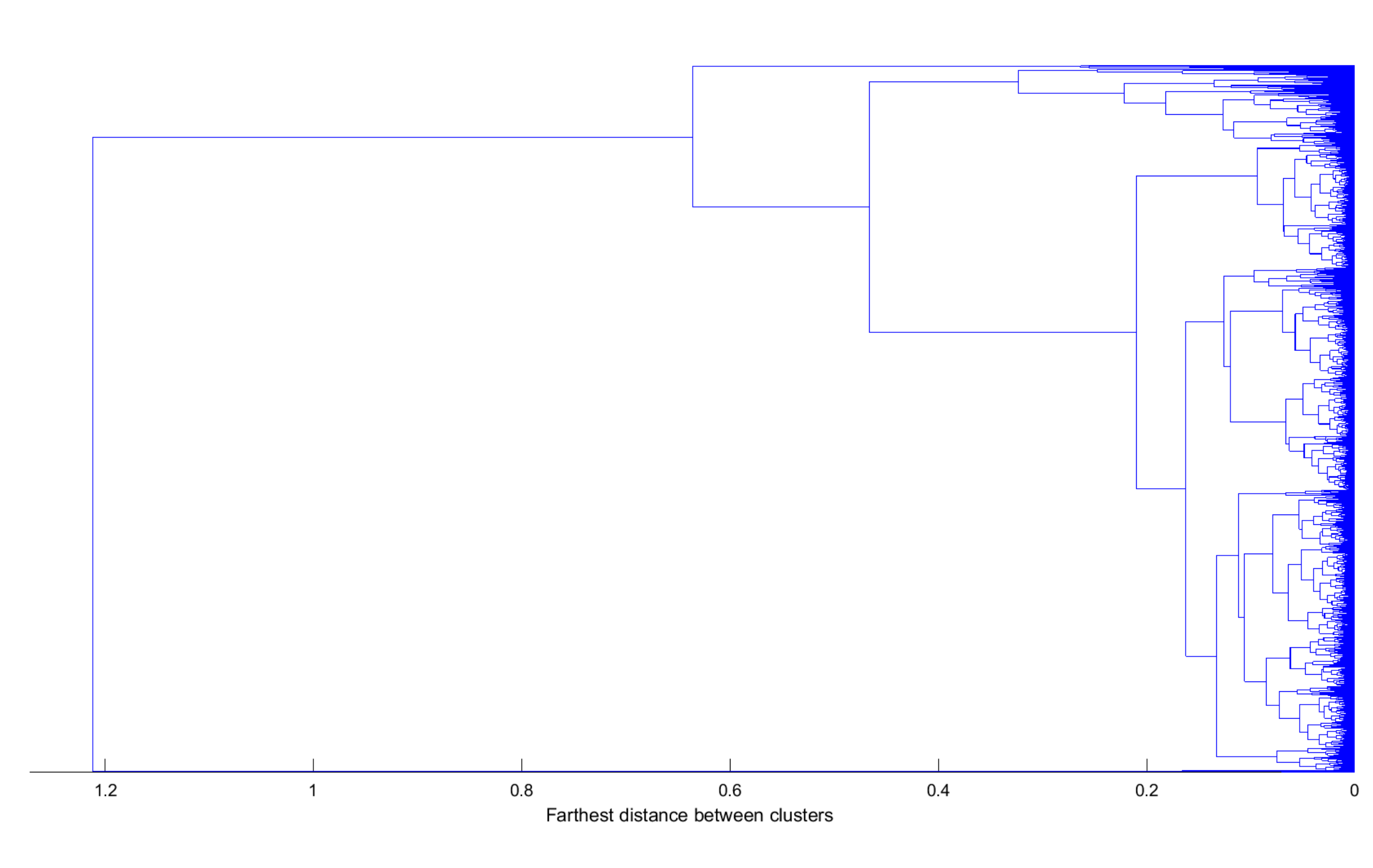}
	\caption{Dendrogram obtained with quantile autocovariance and complete linkage.}\label{dendroQAC}
\end{figure}

\begin{figure}
	\centering
	\includegraphics[width=4.5in]{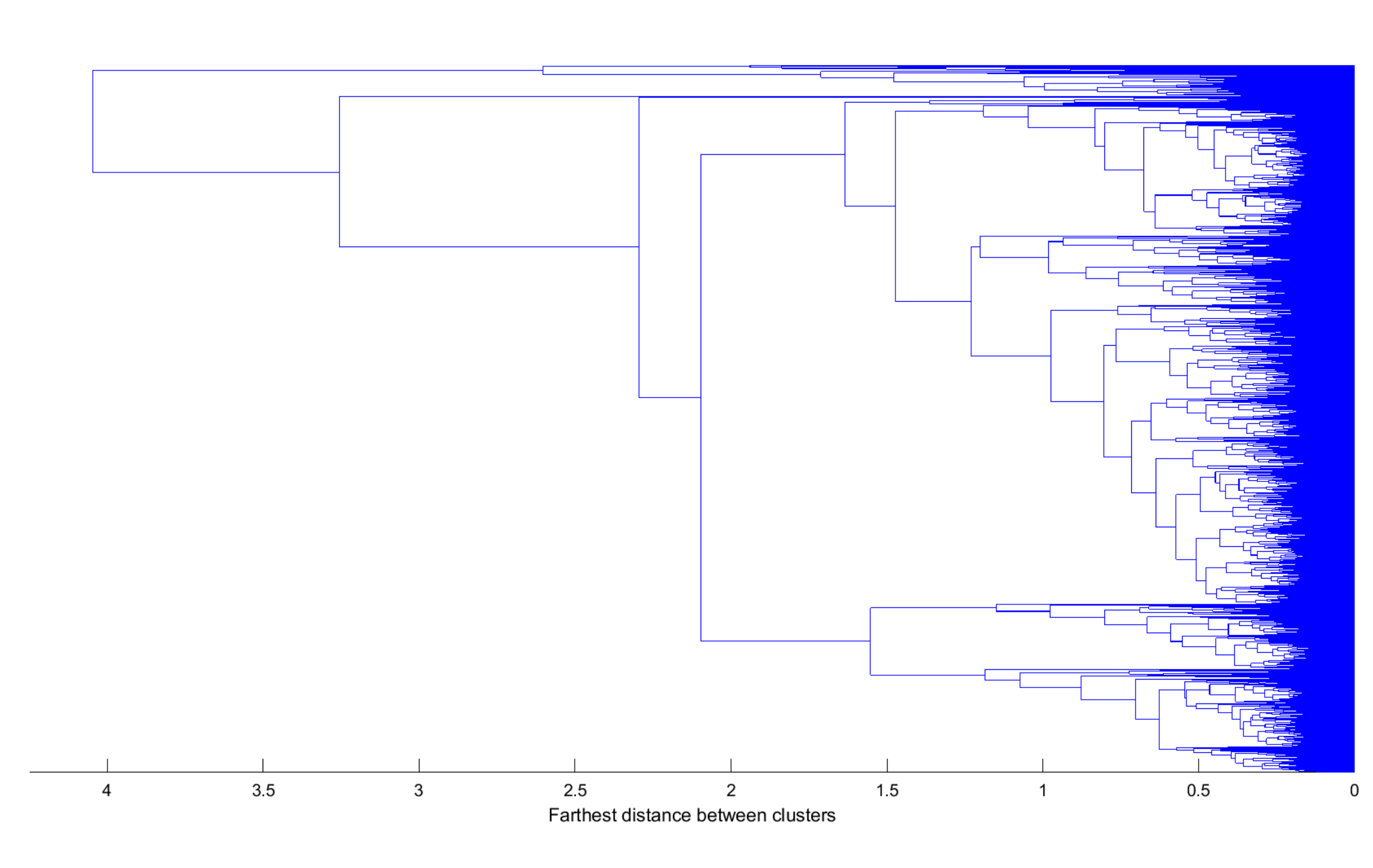}
	\caption{Dendrogram obtained with autocorrelation coefficients and complete linkage.}\label{dendroSAC}
\end{figure}

\begin{figure}
	\centering
	\includegraphics[width=4.5in]{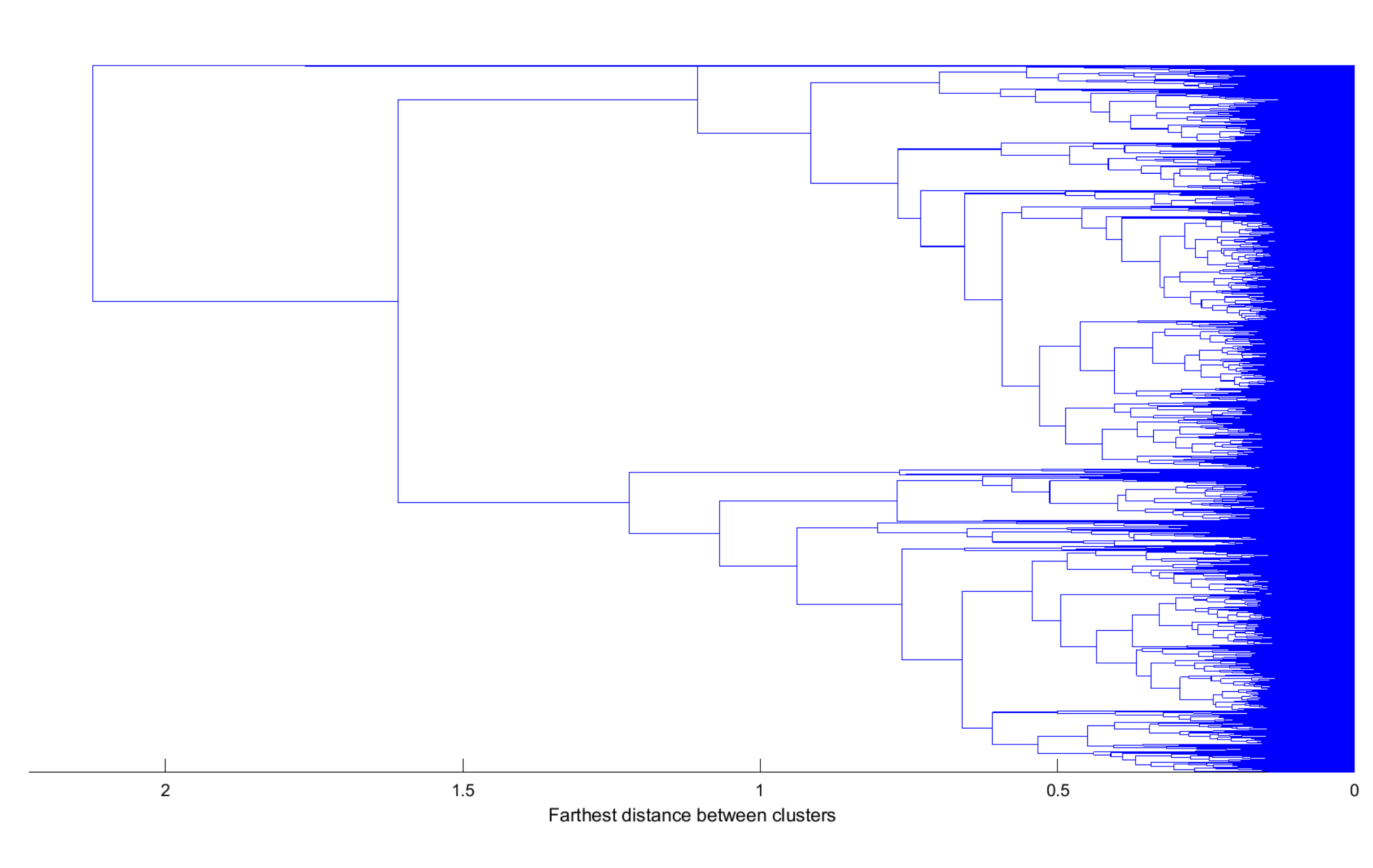}
	\caption{Dendrogram obtained with partial autocorrelation coefficients and complete linkage.}\label{dendroPAC}
\end{figure}

The following three dendrograms, Fig. \ref{dendroQAC} - \ref{dendroPAC}, are obtained using the QC, AC and PAC features and complete linkage introduced in Section \ref{sec:Model}. In the three graphs we can observe some clear groups of observations (time series) and also observations that are joined to the hierarchical structure at large levels. Those observations have a dynamic “atypical” behavior and are grouped in clusters with less than 1\% of the total number of time series. Once we discard the atypical observations, we find eight, six and seven large clusters for QC, AC and PAC, respectively. Moreover, the degrees of coincidence among these three clusters partitions are low as indicated by the adjusted Rand indexes (0.0941 when comparing QC and AC; 0.1432 when comparing QC and PAC and 0.2687 when comparing AC and PAC). This implies that the three approaches look at different characteristics of the time series.

\begin{figure}
	\centering
	\includegraphics[width=4.5in]{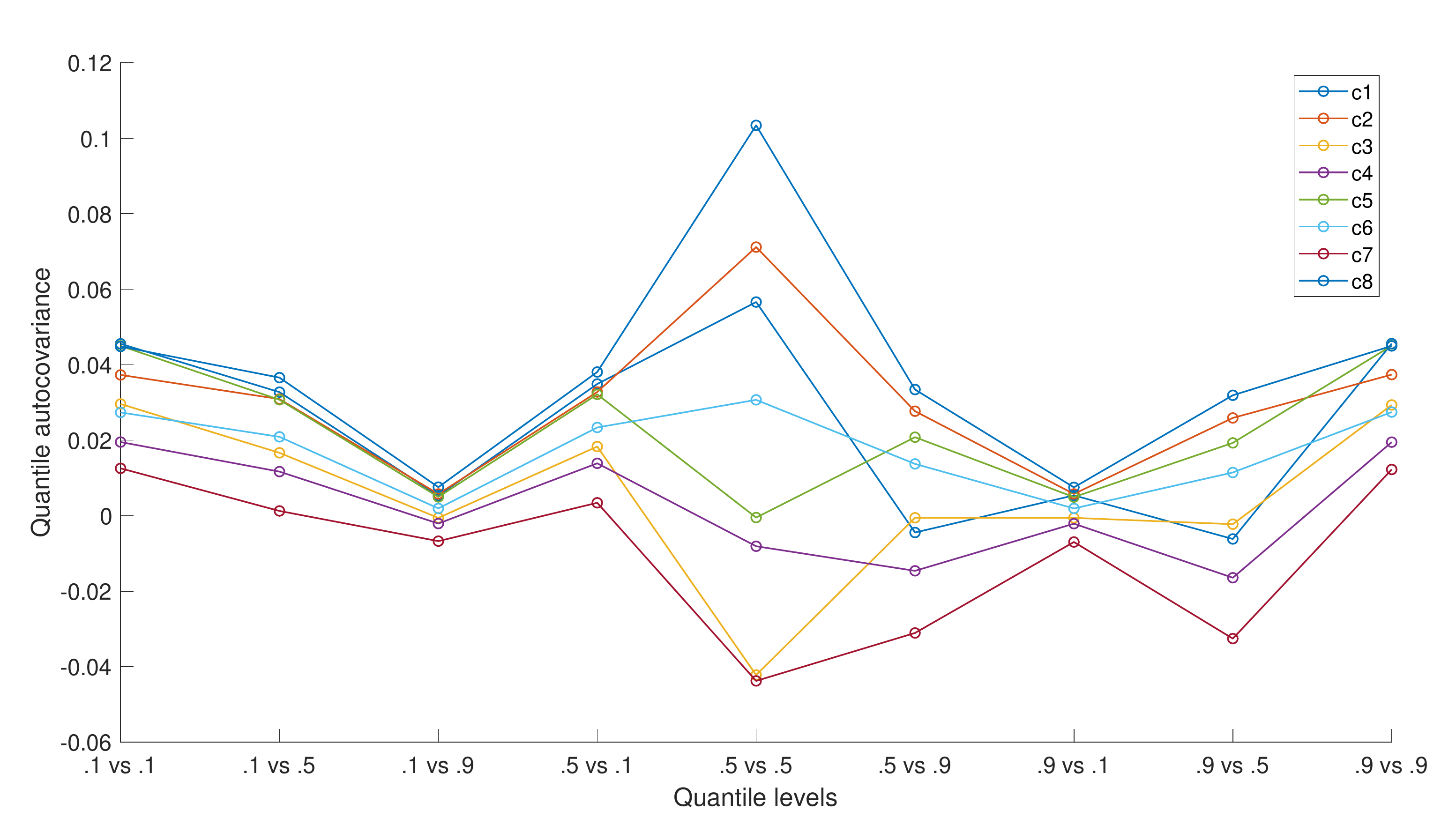}
	\caption{Main clusters obtained with quantile autocovariances and complete linkage.}\label{MclustersQAC}
\end{figure}

\begin{figure}
	\centering
	\includegraphics[width=4.5in]{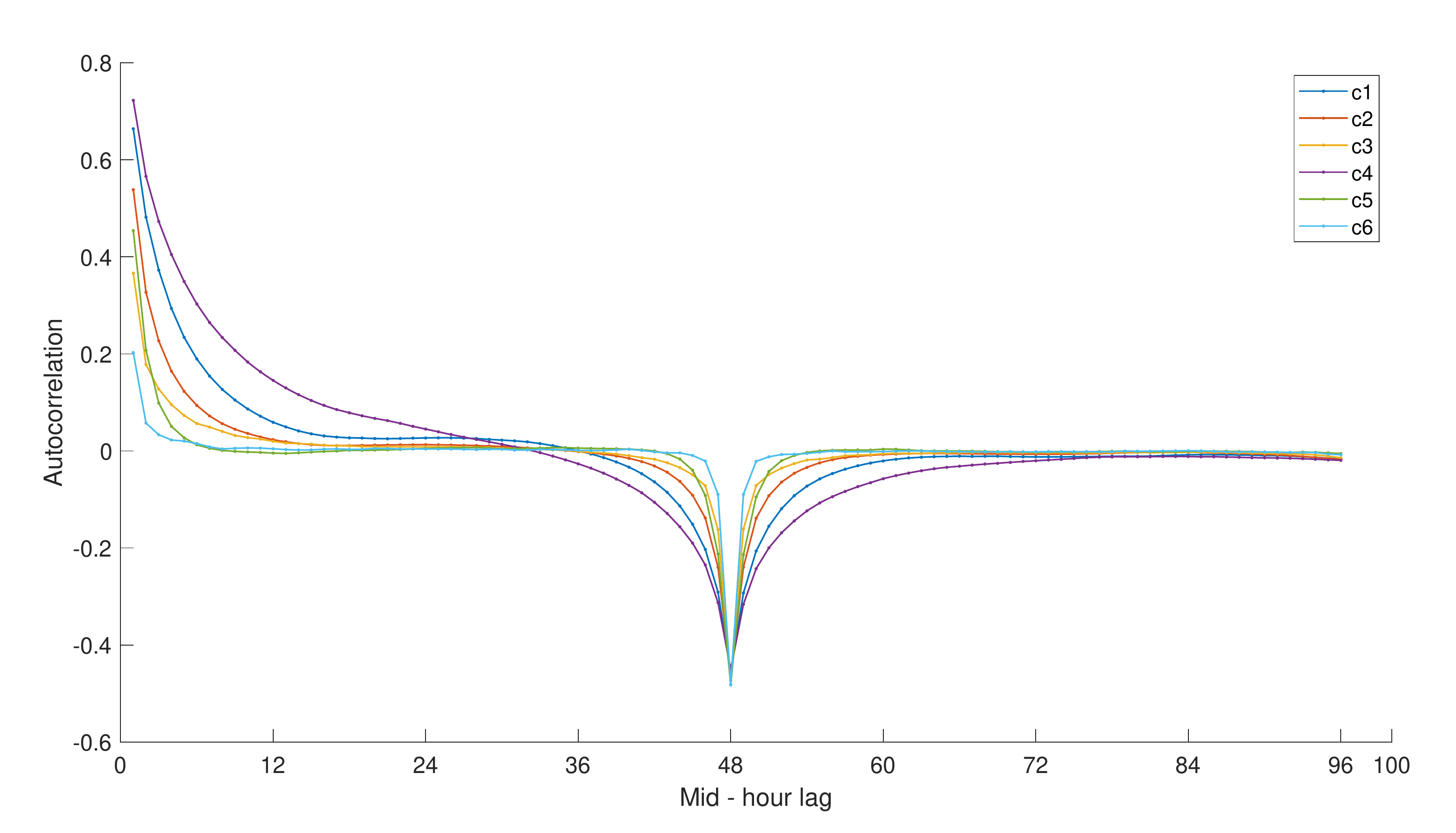}
	\caption{Main clusters obtained with autocorrelation coefficients and complete linkage.}\label{MclustersSAC}
\end{figure}

\begin{figure}
	\centering
	\includegraphics[width=4.5in]{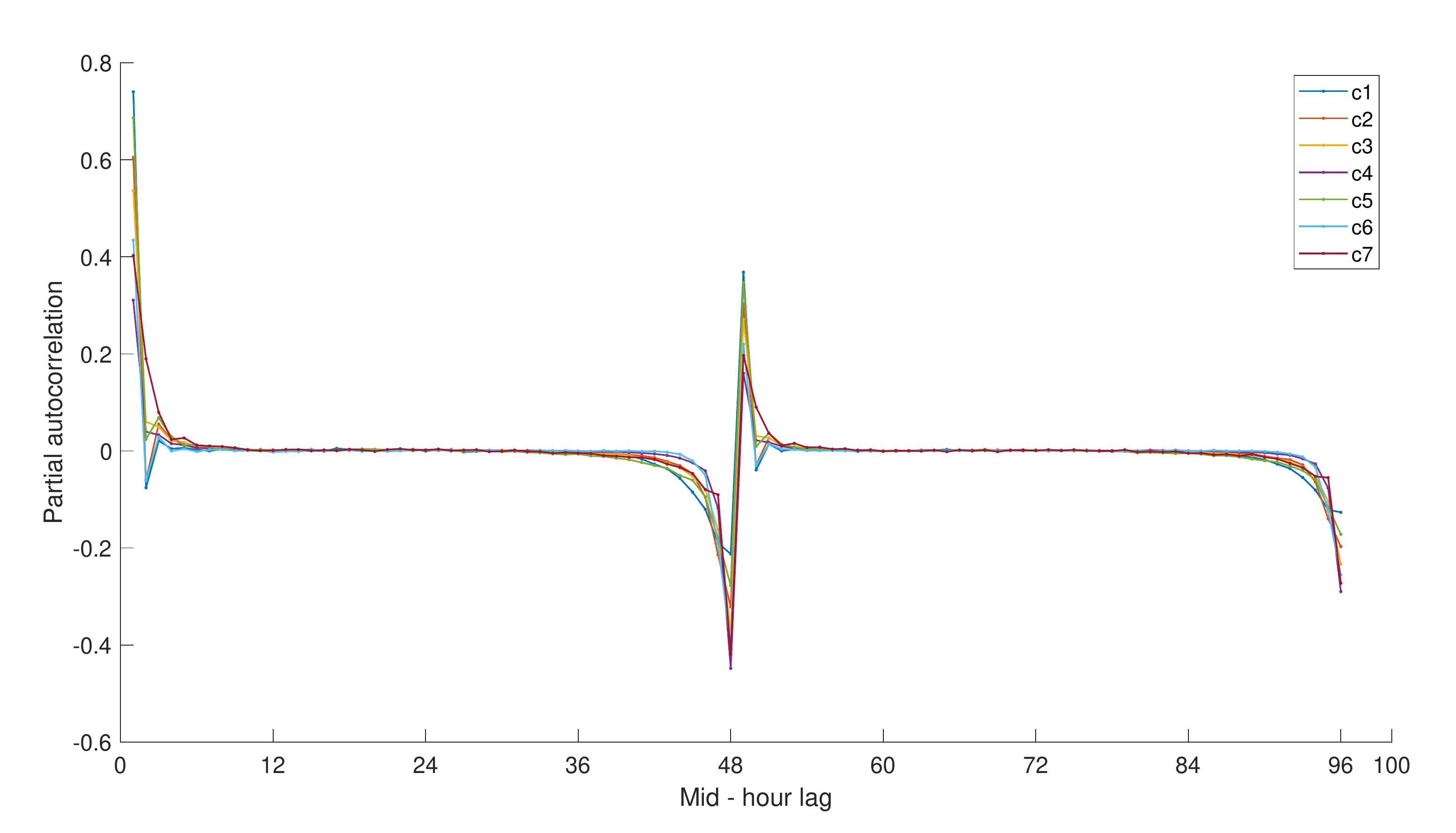}
	\caption{Main clusters obtained with partial autocorrelation coefficients and complete linkage.}\label{MclustersPAC}
\end{figure}

Figures \ref{MclustersQAC} - \ref{MclustersPAC} illustrate these large clusters obtained with QC, AC and PAC, respectively. In the figures, we represent the mean of the features used to obtain the clusters. There are nine features, in the case of QC, corresponding to the covariance of quantiles 10\%, 50\% and 90\%. In the case of AC and PAC, we use the first 96 simple and partial autocorrelations, respectively. The clusters based on QC reveals differences in the median consumptions (.5 versus .5) and highest versus median consumptions (.9 versus .5). For instance, it is remarkable the difference between c3 and c7
versus c1, c2, c4, c6 and c8 at the median consumptions. The c3 and c7 have negative covariances and the c1, c2, c4, c6 and c8 have positive ones. That is,
in the first group, a consumption below the median tends to be followed by consumption above the median, while the second group tends to maintain their
consumption below the median. The groups by SAC and PAC show differences in the short range dependencies but also in the way they are around the lag 48
(one day). We can focus on the first correlations coefficients that show different degrees of persistency in the consumptions. For instance, in the AC
clusters, there is a clear order from high dependency at c4, c1 and c2, medium at c3 and c5 and to low dependency at c6. At the PAC clusters, we can
differentiate between clusters with negative second partial autocorrelation (c1, c2 and c6), medium (c3, c4, and c5) and hight positive (c7). That is,
once we eliminate the first order correlation, there are negative (or positive) direct effects on the consumption at the 2-step ahead period. 

\begin{figure}
	\centering
	\includegraphics[width=4.5in]{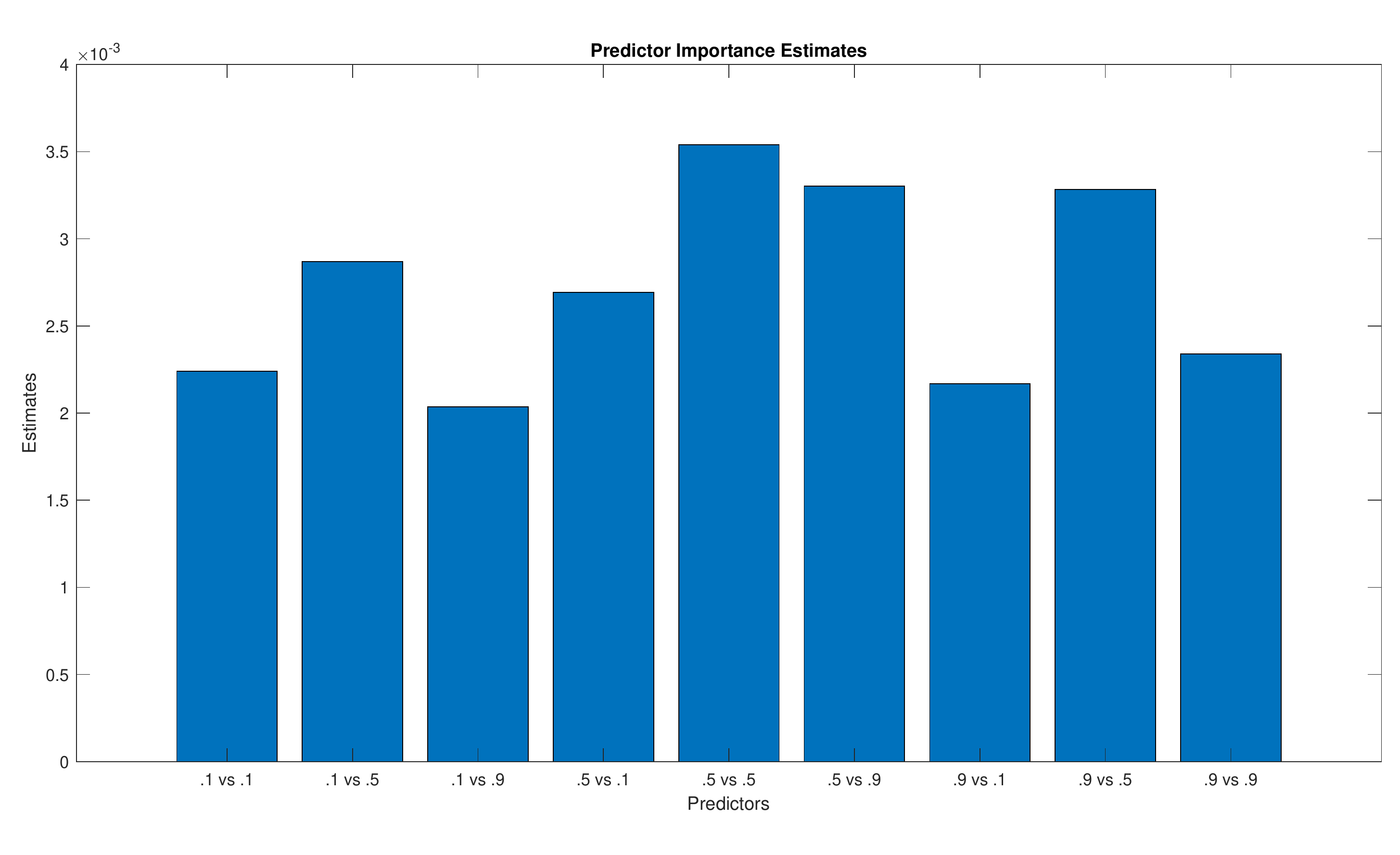}
	\caption{Predictor importance estimates for clusters based on quantile autocovariances.}\label{QACestimates}
\end{figure}

\begin{figure}
	\centering
	\includegraphics[width=4.5in]{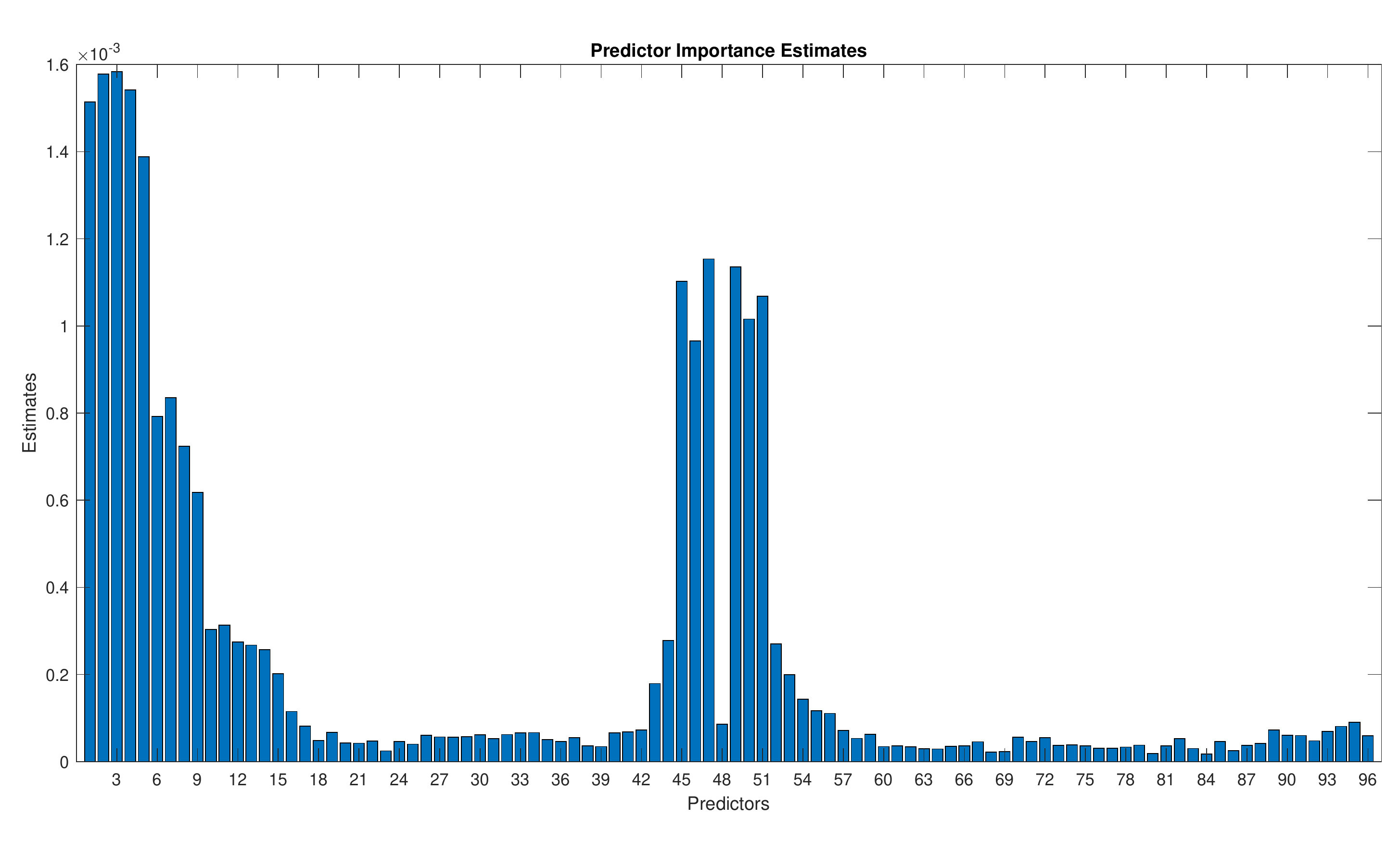}
	\caption{Predictor importance estimates for clusters based on autocorrelation coefficients.}\label{SACestimates}
\end{figure}

\begin{figure}
	\centering
	\includegraphics[width=4.5in]{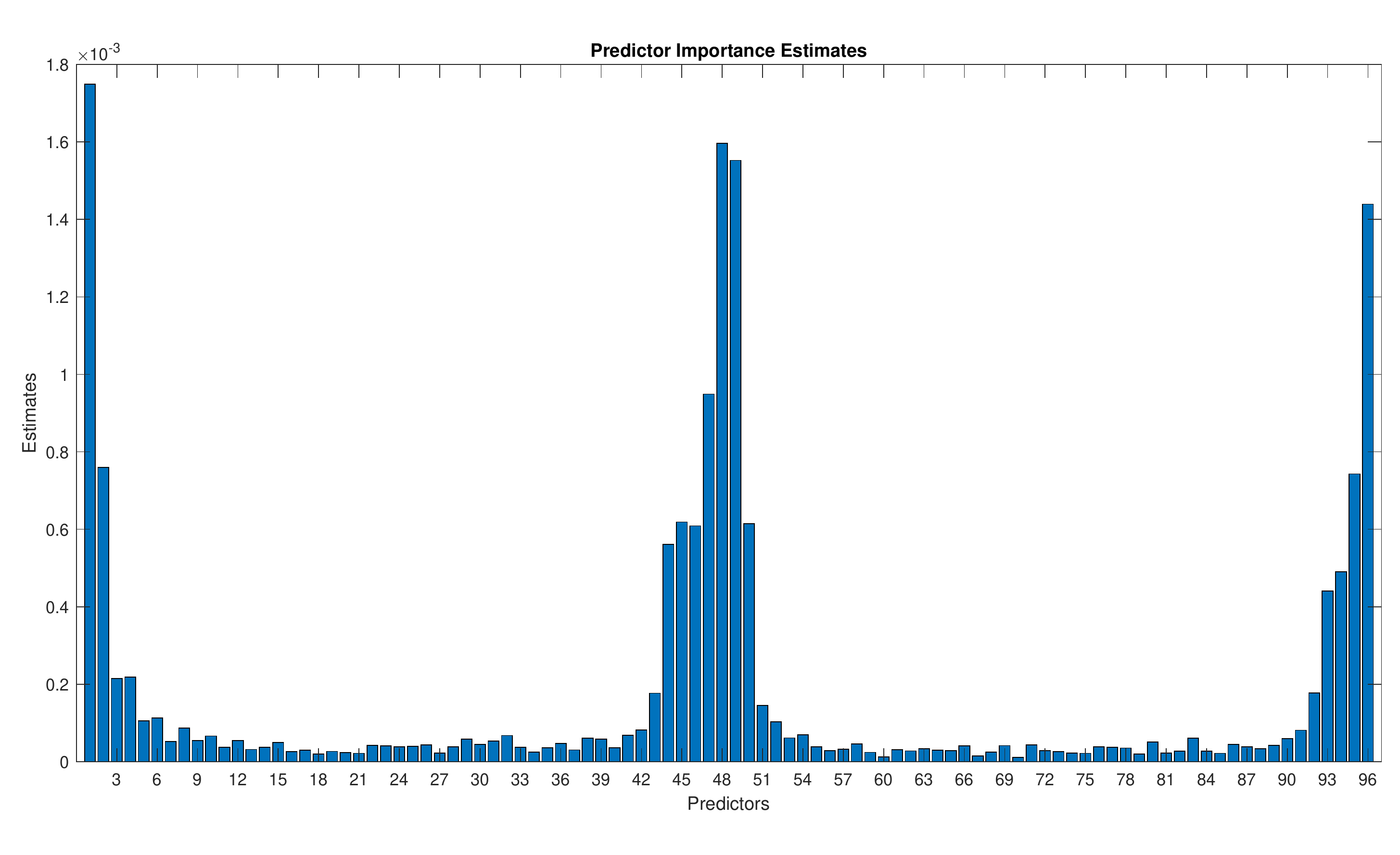}
	\caption{Predictor importance estimates for clusters based on partial autocorrelation coefficients.}\label{PACestimates}
\end{figure}

Figures \ref{QACestimates} - \ref{PACestimates} provide the estimates of the predictor importance. It is clear that all features are relevant in the clustering 
based on QC but we can make a selection of features in the clusterings based on SAC and PAC. In particular, for SAC, the first fifteen lags and the four lags around 
the 48--lag appear to be relevant and, for PAC, the first four lags and the four lags before and two lags after the 48-- and 96--lags as well as those daily 
``seasonal'' lags. It is interesting to notice that the 48--lag is not relevant in the SAC but this is due to the (daily) seasonal difference. However, there are 
still stationary seasonal behavior as reflected by relevant predictors/lags around the 48--lag. For PAC, the daily lags are highly relevant. The misclassification
rates estimated by cross--validation for the three trained decision trees were 9.9\%, 23.3\% and 21.4\% when using QC, SAC and PAC, respectively. These low rates
point out that the obtained trees are good approximations to the clustering mechanism. Of course, other supervised classification procedures such as random forest
or neural networks can be used in order to obtain better approximations.

Tables \ref{QACversusACORNGROUPED} - \ref{PACversusACORNGROUPED} show the number of households on each cluster that are classified in the three ACORN\_GROUPED categories. Note that they are unevenly distributed across clusters. Indeed, we have performed chi-squared tests in those tables and the results are highly significant in the three cases revealing that clustering is related to ACORN\_GROUPED
classification. This shows that the proposed clustering methodology is able to, up to some representative extend, provide insights on the geo-demographic characteristics of a household (Acorn groups), just by studying the time series dependencies.

\begin{table}
	\centering
	\begin{tabular}{c|c|c|c}
		Cluster & Adversity & Comfortable & Affluent \\
		\hline
		c1 &  24 &  24 &  44 \\
		c2 & 293 & 278 & 360 \\
		c3 &  42 &  23 &  36 \\
		c4 &  52 &  40 &  37 \\
		c5 &  28 &  22 &  55 \\
		c6 & 482 & 343 & 358 \\
		c7 &  18 &   9 &  16 \\
		c8 & 146 & 160 & 258 \\
	\end{tabular}
	\caption{Clusters by quantile autocovariance versus ACORN\_GROUPED}\label{QACversusACORNGROUPED}
\end{table}

\begin{table}
	\centering
	\begin{tabular}{c|c|c|c}
		Cluster & Adversity & Comfortable & Affluent \\ \hline
		c1 &  70 &  83 & 161 \\
		c2 & 426 & 404 & 544 \\
		c3 & 252 & 171 & 179 \\
		c4 &  16 &  23 &  59 \\
		c5 & 120 &  79 &  99 \\
		c6 & 214 & 140 & 120 \\
	\end{tabular}
	\caption{Clusters by autocorrelation coefficients versus ACORNGROUPED}\label{SACversusACORNGROUPED}
\end{table}

\begin{table}
	\centering
	\begin{tabular}{c|c|c|c}
		Cluster & Adversity & Comfortable & Affluent \\ \hline
		c1 &  27 &  16 &  61 \\
		c2 &  66 &  67 & 111 \\
		c3 & 397 & 376 & 501 \\
		c4 & 454 & 305 & 261 \\
		c5 &  39 &  51 & 129 \\
		c6 &  83 &  54 &  68 \\
		c7 &  18 &  29 &  39 \\
	\end{tabular}
	\caption{Clusters by partial autocorrelation coefficients versus ACORNGROUPED}\label{PACversusACORNGROUPED}
\end{table}

\begin{figure}
	\centering
	\includegraphics[width=5.5in]{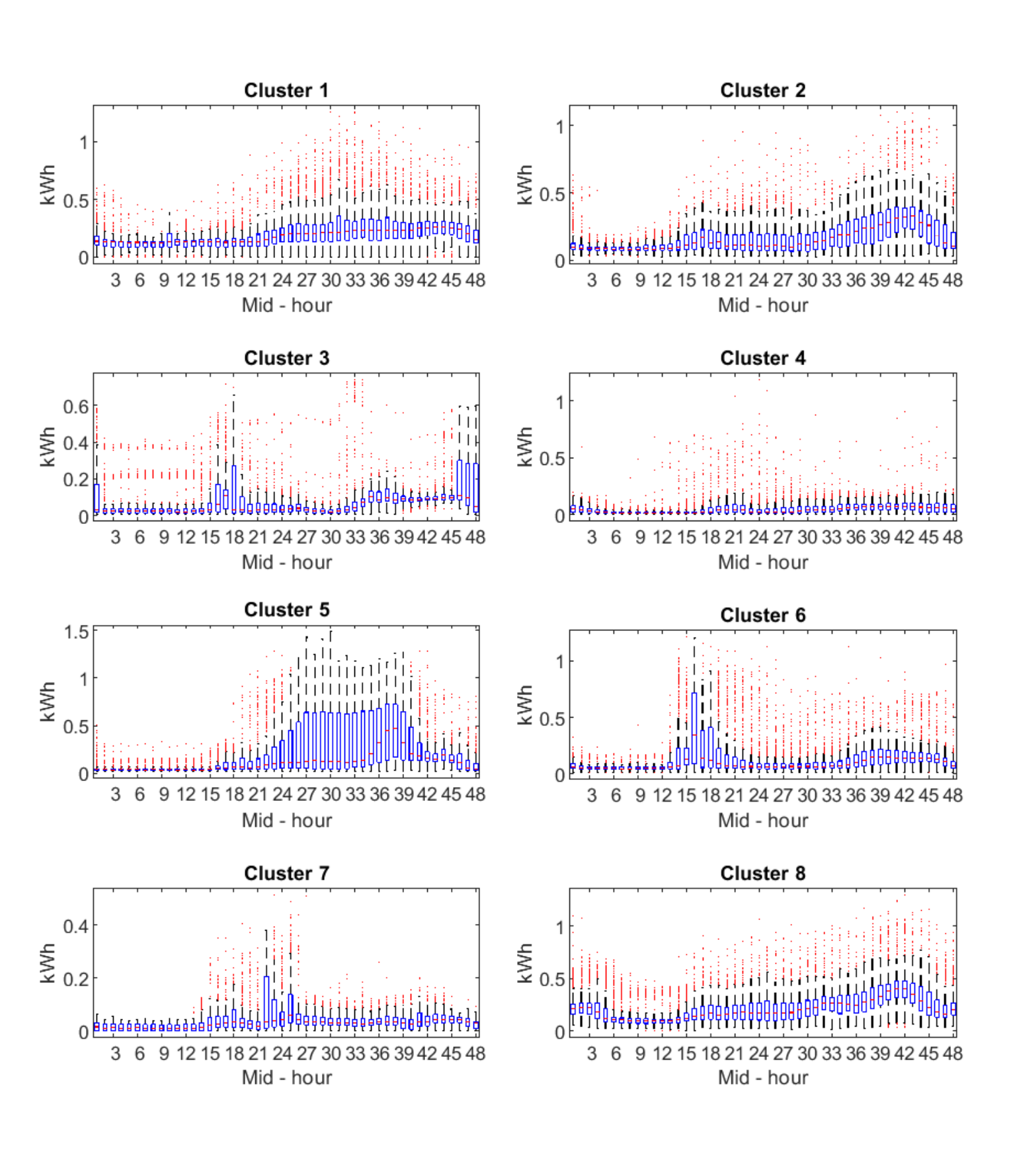}
	\caption{Prototype's hourly profile for clusters obtained with quantile autocovariance and complete linkage.}\label{HourlyQAC}
\end{figure}

\begin{figure}
	\centering
	\includegraphics[width=5.5in]{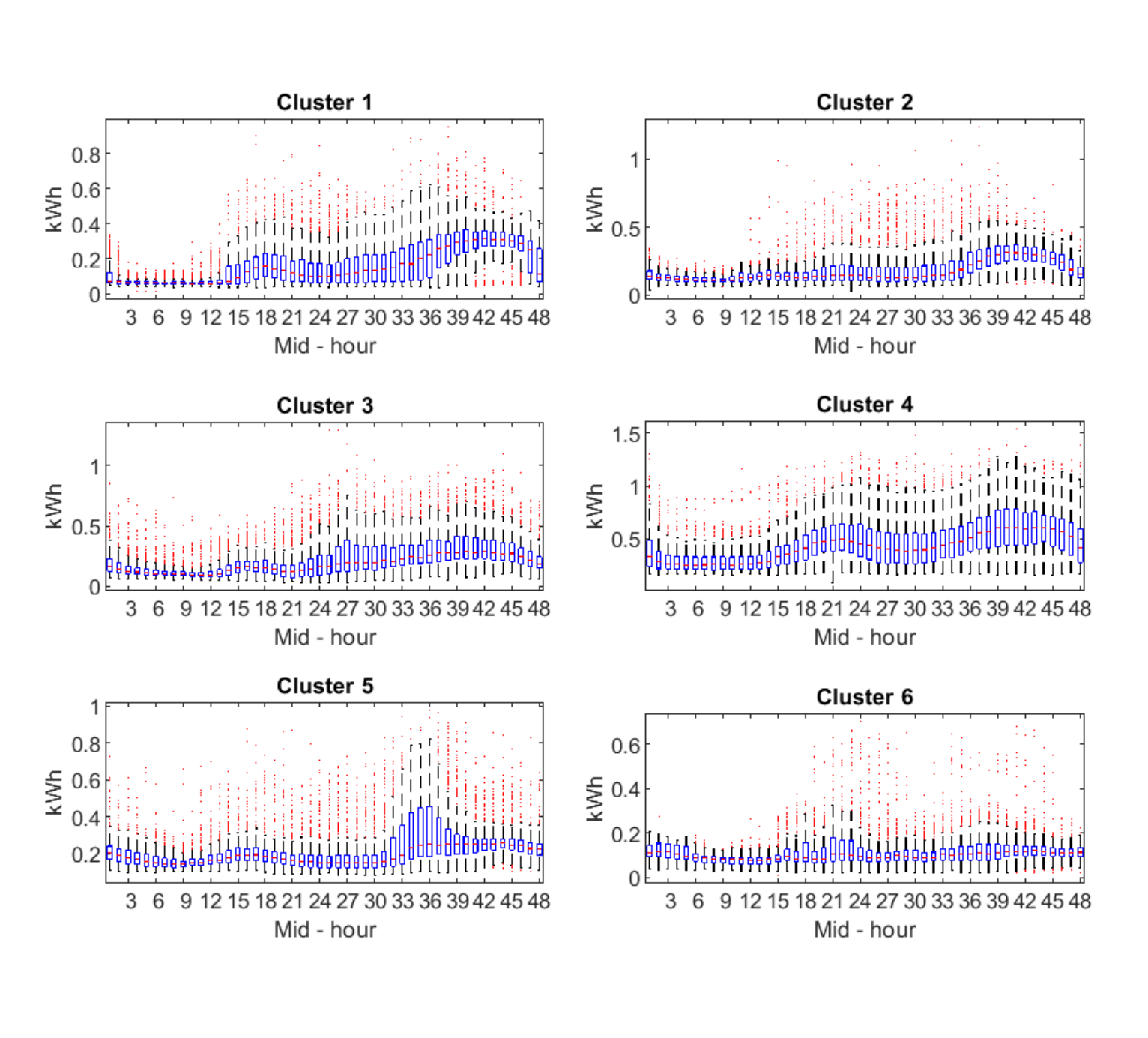}
	\caption{Prototype's hourly profile for clusters obtained with autocorrelation coefficients and complete linkage.}\label{HourlySAC}
\end{figure}

\begin{figure}
	\centering
	\includegraphics[width=5.5in]{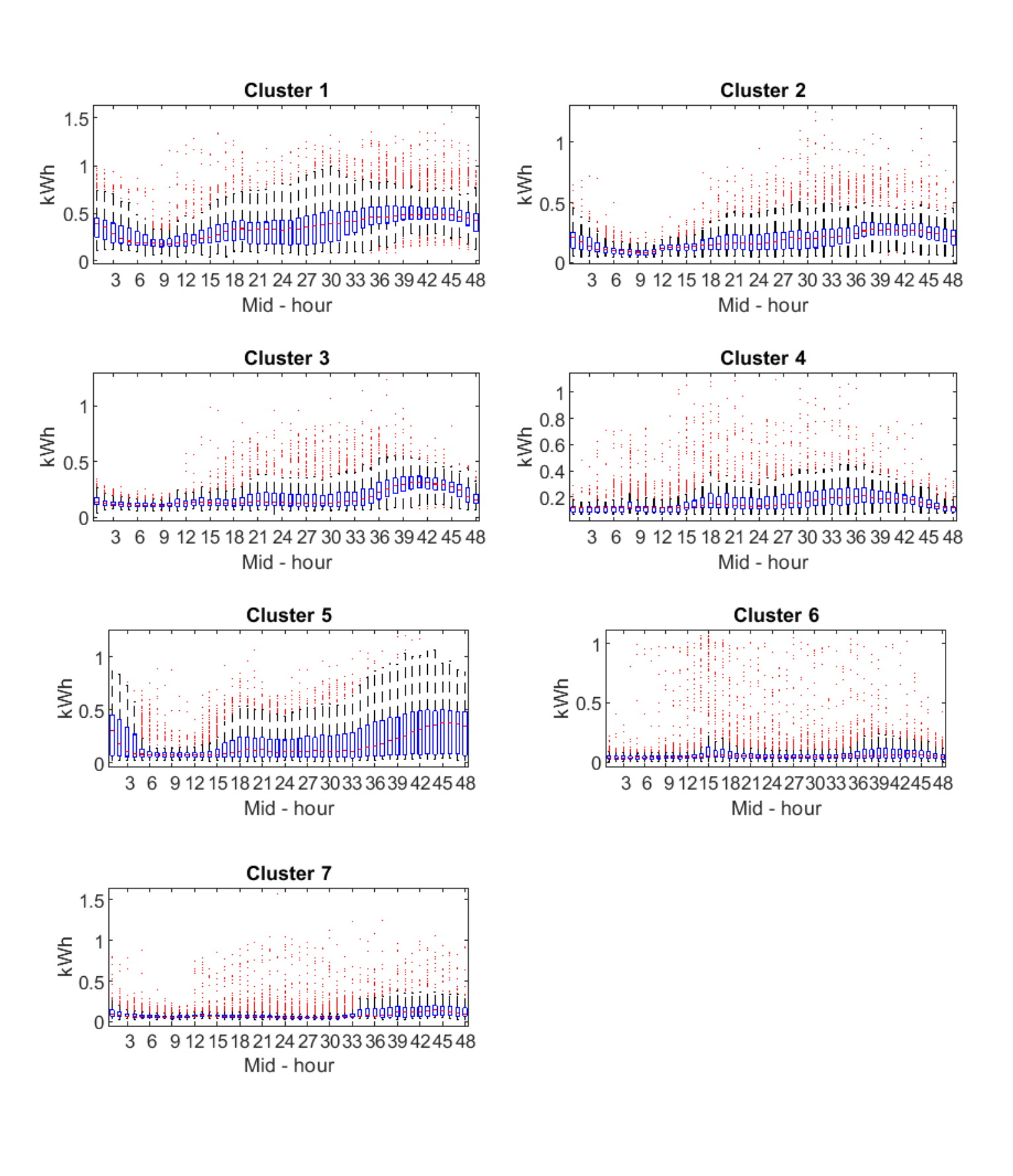}
	\caption{Prototype's hourly profile for clusters obtained with partial autocorrelation coefficients and complete linkage.}\label{HourlyPAC}
\end{figure}

Figures \ref{HourlyQAC} - \ref{HourlyPAC} show the prototype's hourly profile for each cluster. The prototype is the medoid of each cluster, that is, the element in the cluster with minimal average dissimilarity to all objects in the cluster. We can observe different characteristic consumption patterns associated to different types of consumers. For instance, the eight clusters obtained with quantile autocovariance and complete linkage in Figure \ref{HourlyQAC} allow distinguishing between consumers with morning (clusters 3, 6 and 7) and evening (clusters 1, 2, 5 and 8) peak loads, and those with a more constant consumption pattern (cluster 4).
This is also appreciated in the 6 clusters obtained with autocorrelation coefficients and complete linkage in Figure \ref{HourlySAC}. In this case, clusters 1 and 4 capture those consumers  with two intermediate peak loads in the morning and in the evening. Cluster 5 represents consumers with a single peak consumption in the afternoon, and clusters 2, 3 and 5, present consumers with less volatility.

Similarly, Figure \ref{HourlyPAC} shows the clusters obtained with partial autocorrelation coefficients and complete linkage. Clusters 1, 2 and 5 characterize consumers with a steady increasing load that reach its maximum at midnight, while clusters 3, 4, 6 and 7, represent consumers with two intermediate peaks in the morning and evening. In this case clusters from each type are mainly differentiated by the average load consumption levels.

\section{Conclusions}\label{sec:Conclusions}
In this work we have presented three different hierarchical-based clustering strategies based on a set ``dissimilarity" measures computed over: quantile auto-covariances, and simple and partial autocorrelations. The main advantage of this approach is that we can summarize each series in only a set of representative features which makes them very easy to implement (highly efficient), easy to automatize and scalable to hundreds of thousands of series, i.e., valid for real-world applications with large datasets of time series, as the ones obtained from smart meters. We evaluate the performance of these clustering models  with thousands of electricity consumption time series. The results are promising: we are able to obtain highly representative clusters capturing different electricity load consumption patterns and identifying the level of influence of each of the models' features.
Moreover, we have seen how the proposed clustering scheme can provide meaningful insights on the geo-demographic level of a household (Acorn groups), just by analyzing its time series dependencies (autocorrelations).

\section*{Acknowledgment}

The authors gratefully acknowledge the financial support from the Spanish government through project
MTM2017-88979-P and from Fundación Iberdrola through ``Ayudas a la Investigación en Energía y Medio Ambiente
2018''.

\bibliographystyle{IEEEtran}
\bibliography{IEEEabrv,Clustering_one_column}

%

\end{document}